# Electronic transport and device prospects of monolayer molybdenum disulphide grown by chemical vapor deposition

Wenjuan Zhu[1*†], Tony Low[1*], Yi-Hsien Lee[2], Han Wang[1], Damon B. Farmer[1], Jing Kong[3], Fengnian Xia[4+] and Phaedon Avouris[1+]

[1] IBM Thomas J. Watson Research Center, Yorktown heights, NY 10598, USA

[2] National Tsing Hua University, Hsinchu, 30013, Taiwan

[3] Massachusetts Institute of Technology, Boston, MA 02139, USA

[4] Department of Electrical Engineering, Yale University, New Haven, CT 06511, USA

Layered transition metal dichalcogenides display a wide range of attractive physical and chemical properties and are potentially important for various device applications. Here we report the electronic transport and device properties of monolayer molybdenum disulphide ($MoS_2$) grown by chemical vapor deposition (CVD). We show that these devices have the potential to suppress short channel effects and have high critical breakdown electric field. However, our study reveals that the electronic properties of these devices are at present, severely limited by the presence of a significant amount of band tail trapping states. Through capacitance and ac conductance measurements, we systematically quantify the density-of-states and response time of these states. Due to the large amount of trapped charges, the measured effective mobility also leads to a large underestimation of the true band mobility and the potential of the material. Continual engineering efforts on improving the sample quality are needed for its potential applications.

---

[*] These authors contributed equally to the work.
[†] Emails are: W.Z (wenjuan@us.ibm.com), F. X. (fengnian.xia@yale.edu), and P. A. (avouris@us.ibm.com).

## Introduction

Two-dimensional materials are attracting considerable attention due to their unique electronic, optical, and mechanical properties[1]. Following the success of graphene, a group of 2D materials known as the transition metal dichalcogenides (TMD) has begun to garner attention. Among them, molybdenum disulfide ($MoS_2$) is probably one of the most explored TMDs [2,3]. The sizeable (1.8 eV) direct bandgap of monolayer $MoS_2$ [4] makes it a potential material for not only digital electronics, but numerous photonic applications such as light emitter[5], photodetectors[6,7], and solar cells [8]. Excellent mechanical flexibility of $MoS_2$ also makes it a compelling semiconducting material for flexible electronics[9,10]. Most existing studies and device demonstrations were performed on exfoliated $MoS_2$ flakes [11-19]. In particular, field-effect-transistors based on monolayer $MoS_2$ was found to exhibit high on/off ratios of ~$10^8$, steep subthreshold swing of ~70 mV/dec [18,20], with reported electron effective mobility ranging from 1 $cm^2/Vs$ to 480 $cm^2/Vs$ [14,17,20-26] depending on the device structures, dielectric environment and processing[21,27]. These encouraging early reports coupled with continual engineering efforts [28], present a compelling case for monolayer $MoS_2$ as an alternative to traditional organic material or amorphous silicon for low-end applications with basic requirement of an effective mobility of > 30 $cm^2/Vs$ [20] e.g. high resolution displays and photodetection [6,7].

Recently, the advent of mass production technologies has enabled scalable growth of polycrystalline monolayer $MoS_2$ by chemical vapor deposition (CVD) [29-32], hence providing a commercially viable path towards $MoS_2$ electronics at low cost [33]. However, the mobility of CVD $MoS_2$ is typically much lower than its exfoliated counterpart, with reported values in the range of 5 to 22 cm$^2$/Vs [34-36]. The physical origin of the differences between CVD and exfoliated MoS$_2$ are not clear at present, however, structural defects [37] such as vacancies, dislocations, grain boundaries as well as charged interfacial states due to the dielectrics in contact [14] can be responsible for the degradation in mobility. Although this problem has presented a major hurdle to the realization of wafer-scale MoS$_2$ electronics and photonics, systematic studies of it are very few[35]. Very recently, the MoS$_2$ community realized that there technical difficulties to accurate mobility extraction, due to the role of contacts and fringing capacitive contributions [38]. Four-terminal Hall effect measurements are more accurate, but have only been demonstrated at very high carrier densities using a electrolyte gating scheme [26], or at very low temperatures [17]. These Hall measurements were all performed on exfoliated MoS$_2$, and will be more challenging to perform in CVD samples due to significantly larger amount of localized states, as will be discussed in this manuscript.

In this paper, we present a systematic methodology for characterizing electronic properties of scalable CVD MoS$_2$, reporting also the presence of significant amounts of band tail states and their profound impact on the electrical device performance. The density distribution and dynamics of these trap states of CVD MoS$_2$ are characterized through systematic capacitance and ac conductance measurements. Extraction of basic electronic transport quantities like the mobility edge and effective mobility are performed using four-probe current measurements. Complementary modeling allows us to draw insights into relevant device quantities such as the fractional occupation of band and trap states, band mobility and the anomalous sub-threshold slope. Lastly, high field electrical behavior such as drain-induced barrier lowering and critical breakdown fields are examined.

## Results

**Characterization of monolayer CVD MoS$_2$**

Monolayer MoS$_2$ were synthesized by CVD, using solid sulfur (S) and molybdenum oxide (MoO$_3$) as the precursors and perylene-3,4,9,10-tetracarboxylic acid tetra-potassium salt (PTAS) [39] as the seed for the CVD growth (see Methods for details). Hall-bar devices were fabricated on these CVD MoS$_2$ mono-layers. Fig. 1a illustrated the schematic of the device. The top gate dielectrics are 2nm aluminum deposited by e-beam evaporation and re-oxidized as a seed layer, followed by 30nm HfO$_2$ deposited by atomic layer deposition (ALD). The details of the device process are discussed in Methods. The atomic force microscopy (AFM) image and the step height profile at the boundary of a MoS$_2$ triangular area are shown in Fig. 1b and the inset of Fig. 1b, respectively. The thickness of the MoS$_2$ layer is measured to be about 0.8 nm, confirming its monolayer character. The Raman spectrum of the CVD MoS$_2$ is shown in supplementary Fig. S1. The E$_{2g}$ and A$_{1g}$ modes are at around 383 and 403 cm$^{-1}$, respectively. Comparing the peak position to the spectrum obtained from the exfoliated monolayer MoS$_2$ [40], we further verify that the MoS$_2$ film is monolayer.

**Density and dynamics of band tail states**

At present, the reported values of the electron mobility in CVD-grown MoS$_2$ devices [34-36] are at least two orders of magnitude smaller than the intrinsic limit [41], suggesting a high degree of disorder and scattering. An inhomogeneous potential distribution in a semiconductor leads to the smearing of the band-edge and the formation of a tail of band-gap state [42]. For example, this inhomogeneity could be the result of a random distribution of trapped charges in sulfur vacancies in MoS$_2$ itself [43], or at MoS$_2$–dielectric (SiO$_2$ or high-k dielectric) interfaces [14]. Structural defects [37], e.g. simple vacancies [43], dislocations and grain boundaries would also lead to localized gap states.

The electronic states in the band tail can be characterized using standard capacitance and ac conductance measurement commonly used in the study of semiconductor devices [44]. Here, these localized states respond like traps with different time constants $\tau_{it}$, and are electrically equivalent to an additional capacitance and resistance in parallel to the semiconductor capacitance. The gate-to-channel capacitance and resistance were measured on the Hall-bars, with high terminal on the top gate and low terminal on the source, drain and all four voltage probing leads simultaneously, as shown in supplementary Fig. S2. The measured capacitance as a function of frequency is shown in Fig. 2a. The observed double hump feature indicates at least two types of traps with different time constants at a given gate voltage. Herein, we denote these traps by the labels "M" and "B", for reasons which will be made apparent below. The equivalent circuit model of the device is shown in Fig. 2b with total impedance given by:

$$Z = (Y_{itB} + Y_{itM} + i\omega C_s + i\omega C_j)^{-1} + (i\omega C_{ox})^{-1} + r_s \quad (1)$$

where ω is the angular frequency, $C_s$ is the quantum capacitance of the MoS$_2$, $C_j$ is the parasitic capacitance, $C_{ox}$ is the oxide capacitance, $r_s$ is the series resistance, $Y_{itM}$ and $Y_{itB}$ are the traps' admittance. Here $Y_{itB} = [\tau_{itB}/C_{itB} + 1/(i\omega C_{itB})]^{-1}$, where $C_{itB}$ and $\tau_{itB}$ are the capacitance and time constant of trap B. The trap capacitance $C_{itB}$ is related to the trap density $D_{itB}$ via: $C_{itB} = eD_{itB}$, where $e$ is the elementary electric charge. Similar expressions apply to trap M. The measured capacitance in series mode $C_{ms}$ is related to the imaginary part of the total impedance Z by:

$$C_{ms} = -\frac{1}{\omega * \text{Im}[Z]} \qquad (2).$$

The lines in Fig. 2a are the fits using this model. We can see that the model provides an excellent fit to the experimental data. From the fitting, we can extract the density of traps and their time constant as a function of gate voltage, shown in Fig. 2c. Here, we observe that traps "M" and "B" are populated predominately within the "mid-gap" and "band-edge" regions, respectively.

Alternatively, the density and time constant of the traps can be extracted from the ac conductance $G_p$ [44]. The ac conductance is obtained from the measured capacitance and resistance. The measured resistances as a function of gate voltage at various frequencies are shown in supplementary Fig. S3. The extraction method of ac conductance are described in more detail in the Supplementary Note 1. The extracted ac conductance over angular frequency $G_p/\omega$ is plotted as function of the driving frequencies $f$ in Fig. 2d. Since the density-of-states of type B traps is several orders of magnitude larger than that of type M in this gate bias range, we expect the former to dominate the ac conductance. The relation between $G_p(\omega)$ and the trap density $D_{itB}$ is given as [44]:

$$\frac{G_p(\omega)}{\omega} = \frac{eD_{itB}}{2\omega\tau_{itB}}\ln[1+(\omega\tau_{itB})^2] \qquad (3)$$

from which one can deduce the $D_{itB}$ and the respective time constant $\tau_{itB}$ from the following simple relations [44]:

$$D_{itB} = \frac{2.5}{e}\left(\frac{G_p}{\omega}\right)_{peak} \tag{4}$$

$$\tau_{itB} = \frac{1.98}{2\times 3.14 \times f_0} \tag{5}$$

Here, $(G_p/\omega)_{peak}$ is the maximum $G_p(\omega)/\omega$ value, and $f_0$ is the frequency at which this maximum is obtained. Repeating the above procedure for different top gate voltages $V_{TG}$, allows us to extract $D_{itB}(V_{TG})$ and $\tau_{itB}(V_{TG})$ as shown in Fig. 2c. From Fig. 2c, we see that the density-of-states and time constant of B type traps extracted based on the capacitance and ac conductance are in good agreement.

Below, we present parameterized models for the electronic density-of-states and their time constants fitted to the experiments. We describe the electronic density-of-states of the measured distributed trap states at the band-edge (i.e. type B) and the extended states with two piece-wise functions as follows:

$$D_n(E) = \begin{cases} \alpha D_0 \exp\left[\dfrac{E-E_D}{\varphi}\right] + D_{itM} & , \quad E_D - \tfrac{1}{2}E_G < E < E_D \\ D_0 - (1-\alpha)D_0 \exp\left[-\dfrac{E-E_D}{\varphi'}\right] + D_{itM} & , \quad E > E_D \end{cases} \tag{6}$$

and illustrate them in Fig. 2c and 2e. Here, $D_0$ is the 2D density-of-states for perfect crystalline MoS$_2$, taken to be $3.3\times 10^{14}$ eV$^{-1}$cm$^{-2}$, consistent with an effective mass of $0.4\,m_0$ at the conduction band minimum at the K valley for monolayer MoS$_2$ [45]. Here we ignored contribution from the satellite valley along ΓK with energy 200meV higher than the conduction band minimum [41,46], because the carrier population is insignificant at these energies at our biasing range. Furthermore, our MoS$_2$ devices are typically n-type doped. $\varphi$ is the characteristic energy width of the band tail. In the limit of $\varphi = 0$, $D_n(E)$ becomes a step function as required for perfect 2D crystals with the conduction band edge situated at $E_D$. $\varphi'$ is chosen so that the two piece-wise functions have continuous gradients at $E_D$. Solving the electrostatics problem, to be described below, a best-fit to the experimentally extracted density-of-states yields the parameter set: $\alpha = 0.33$ and $\varphi = 100 meV$. The comparison between the model $D_n(V_{TG})$ and the measured band tail states $D_{itB}(V_{TG})$ is shown in Fig. 2c, and the mid-gap states $D_{itM}(V_{TG})$ is described by an error function instead. The traps' response time are fitted to an exponential model,

$$\tau = \tau_0 \exp\left[-\frac{E - E_{mid}}{\Phi_\tau}\right] \quad (7)$$

and the comparison with experimental data is shown in Fig. 2c.

With the parameterized density-of-states model, we can calculate the ac capacitance and compare against experiments. The Poisson equation describing the electrostatics of the problem can be expressed as:

$$Q_n + Q_0 = \varepsilon_{ox}\left(\frac{V_{TG} + E_D/e}{t_{top}}\right) \quad (8)$$

where $t_{top}$ is the "effective-oxide-thickness" for the top gate dielectric, $\varepsilon_{ox}$ is the dielectric constant of silicon oxide, $Q_0$ is a constant which includes contributions from the fixed charges and doping in as-prepared MoS$_2$ etc, and $Q_n$ is the electronic charges in the smeared out conduction band and can be computed from $Q_n = e \int D_n(E) f_n(E, E_F) dE$ where $f_n$ is the Fermi Dirac function. In solving for the self-consistent electrostatics described by Eq. 8, the Fermi energy $E_F$ is taken to be the reference i.e. $E_F = 0$. Once the electrostatics is determined, the admittance associated with each of the traps can then be computed via,

$$Y_{it} = e^2 \int D_{it} \frac{j\omega + \omega^2 \tau_{it}}{1+(\tau_{it}\omega)^2} \frac{\partial f_n}{\partial E} dE \tag{9}$$

and the total capacitance can be computed employing the equivalent circuit model in Fig. 2b, and using equation (1) and (2). Reasonable agreement with the measured ac capacitance is obtained as shown in Fig. 2f.

To summarize, the electronic density-of-states model described above is well-calibrated to the experimentally measured density and dynamics of the band tail states. It follows an exponentially decaying behavior, with a significantly large energy width of $\varphi = 100 meV$, suggesting a high degree of potential disorder and scattering. This model will be employed in the subsequent discussion to obtain other quantities of interest, such as the band mobility.

**Mobility edge**

The concept of a "mobility edge" has greatly facilitated our understanding of electronic transport in a disordered system [47]. The mobility edge is a boundary located in the band tail, in which states above it are extended states with band transport, while those below it are localized states that conduct via thermally assisted mechanisms such as Mott variable range hopping (VRH) [48-50] or an Arrhenius-type activated behavior [51]. Four point variable temperature measurements were performed on our devices from 4.4 to 400 K as shown in Fig. 3a. We found that neither the VRH nor the Arrhenius model can individually describe the data satisfactorily over the whole temperature range. It is very likely that a combination of both transport mechanisms might be operating here. For example, the VRH usually dominates for localized states deep in the band tail, while the Arrhenius-type activated behavior is more likely for shallow localized states.

The conductance versus the inverse of temperature ($1/T$) is shown in Fig. 3b, showing the exponential decrease with $1/T$ over the intermediate temperature range, where the conductance $G$ can be described by:

$$G = G_0 e^{-E_a/k_B T} \tag{10}$$

where $E_a$ is the activation energy, $k_B$ is the Boltzmann constant and $G_0$ is a fitting parameter. This allows us to extract the activation energy $E_a = E_M - E_F$. The measured $E_a$ (versus $V_{TG}$) is compared to the model (see Fig 3b inset), which allows us to determine the location of the mobility edge $E_M$ in the energy band picture. We found that in our devices $E_M$ is $\approx 0.01 eV$ above $E_D$, as illustrated in Fig. 2e. We also noticed a departure from activated behavior at lower temperature or bias, indicating the possible onset of an additional transport mechanism such as VRH. The extracted activation energy in the inset of Fig. 3b suggests that in most of our gate biases, the Fermi level does not exceed the mobility edge energy. However, at large $V_{TG}$ i.e. $V_{TG} > 2V$, the Fermi energy is within tens of meV from the mobility edge. Hence, one can expect an appreciable fraction of extended states occupation due to thermal smearing. The calculated band carrier density as a function of gate voltage at various temperatures are shown in Fig. 3c.

**Transport coefficients**

The "effective mobility", or sometimes referred to as the "drift mobility", is a commonly used transport coefficient in semiconductors [52]. It is defined as the ratio of the measured conductivity to the total charge density i.e. $\mu_{eff} = \sigma/Q_{total}$. The total charge density $Q_{total}$ is typically estimated from $C_{ox}(V_g - V_T)$, where $C_{ox}$ is the oxide capacitance, $V_g$ is the gate voltage, and $V_T$ is the threshold voltage. We extract $C_{ox}$ from the measured capacitance at strong accumulation. Fig. 4a shows the effective mobility as a function of gate voltage at different temperatures. The effective mobility is <10 cm$^2$/V-s over the range of temperature and applied bias in our experiments. Contact resistance is eliminated in our measurement which employs a four-probe scheme. The measured effective mobility is significantly lower than the phonon-limited intrinsic mobility in monolayer MoS$_2$, predicted to be over 400 cm$^2$/V-s [41], and the highest measured mobility of ~200 cm$^2$/V-s in exfoliated MoS$_2$ devices[20]. However, it is consistent with results on similar CVD-grown MoS$_2$ devices which report mobilities in the range of 5 to 25 cm$^2$/V-s [35]. The significantly lower mobility for CVD-grown MoS$_2$, is to a large part, due to the presence of traps.

The total charge density includes both the free and trapped charges: $Q_{total} = e(n_{loc} + n_{band})$ where $n_{loc}$ and $n_{band}$ refer to the density of occupied states below and above the mobility edge, respectively. Another commonly used transport coefficient in a disordered system is the "band mobility" [53,54]. It is defined as the ratio of the measured conductivity to the density of occupied states above the mobility edge (i.e. extended states):

$$\mu_{band} = \frac{\sigma}{en_{band}} = \mu_{eff} \frac{n_{loc} + n_{band}}{n_{band}} \tag{11}$$

In general, the density of the extended states is difficult to measure, since the large amount of localized states would result in large noise-to-signal ratio the in the Hall measurement. Up to now, the Hall effect has only been observed in exfoliated MoS$_2$ with very high carrier density induced by an electrolyte gating scheme [26] or at very low temperatures [17].

Previously, we obtained $D_n(E)$ in conjunction with $E_M$ in the energy band picture. Solving the electrostatics in conjunction with the above information would allow us to estimate the fraction of localized and extended states. Fig. 3c plots the computed $n_{band}(V_{TG})$ at different temperatures. The result indicates that $n_{band}$ only accounts for less than 25% of $n_{total}$. Only at high temperature or bias, $n_{band}$ can exceed 25% i.e. $T > 300K$ and $V_{TG} > 2V$. The comparison between the extracted band mobility and measured effective mobility is shown in Fig. 4b. The band mobility is several times higher than the effective mobility, but still significantly lower than the phonon-limited mobility as predicted in Ref. [41]. This mobility degradation may involve many sources of scattering. For example, structural defects in CVD MoS$_2$ layer and grain boundaries can induce short-range scattering. Surface polar phonon either in the high-k dielectrics (HfO$_2$ and AlO$_x$) or in the SiO$_2$ substrate underneath can also play a role. However, the significant trap population measured here suggests that Coulomb scattering due to trapped charges is the likely limiting factor for the electron mobility. Due to the parabolic band-structure of MoS$_2$, the energy averaged scattering time due to Coulomb scattering should increase proportionally to temperature[55] $\mu \propto k_B T$. This is also consistent with the observed trend in Fig. 4b for temperature below 300K. Above this temperature phonon scattering takes over.

**Sub-threshold swing**

In an ideal semiconductor, the sub-threshold swing is given by [52]: $S = k_B T \ln(10)(1 + C_D / C_{ox})$, where $C_D$ is the depletion capacitance. The sub-threshold swing increases linearly with temperature, since the carrier density increases exponentially with temperature $n \propto \exp(\frac{E_F - E_i}{k_B T})$. In our device, we observed that the sub-threshold swing is ≈200mV/dec and nearly independent of temperature, as shown in Fig. 3a. Similar observations were made on MoS$_2$ flakes [11]. This departure from the ideal behavior can be understood by recalling that the band tail is distributed in energy, i.e. $\exp\left[\frac{E - E_D}{\varphi}\right]$, with an energy width that is significantly larger than the thermal energy i.e. $\varphi \gg k_B T$. Indeed, the calculated sub-threshold behavior confirms that temperature does not have a significant effect, as shown in Fig. 3c. Hence, the observed temperature independent sub-threshold swing reinforced our earlier conclusions on the existence of band tail states.

**Drain-induced-barrier-lowering**

The electrostatic integrity of an electronic device upon downscaling is often quantified by evaluating the amount of drain-induced-barrier-lowering (DIBL) [56]. This measures the reduction in threshold voltage due to the applied drain bias. A common approach used to suppress DIBL involves reducing the channel thickness, since the minimum channel length needed to preserve the long channel behavior is typically ~4-5 times the electrostatic scaling length $\lambda = \sqrt{\varepsilon_s t_s t_{ox}/\varepsilon_{ox}}$ for a planar device structure[57,58], where $\varepsilon_s$ and $\varepsilon_{ox}$ are the dielectric constants of the semiconductor and the gate oxide and $t_s$ and $t_{ox}$ are the thicknesses of the semiconductor and gate oxide, respectively. In this regard, thinner silicon has been pursued by using SOI (silicon-on-insulator) and ETSOI (extremely thin silicon-on-insulator). However, the mobility degrades dramatically as the thickness is scaled down due to surface roughness [59,60]. Atomically thin 2D semiconducting material such as $MoS_2$ and $WSe_2$ are promising candidates in this regard. The typical DC performances of $MoS_2$ MOSFETs with various channel lengths are shown in Supplemental Figure S4. Fig. 5a shows the DIBL of CVD $MoS_2$ MOSFET with variable channel lengths from 4μm to 32nm. Despite the thick dielectric used in our MOSFETs, (~34nm $HfO_2$/ $AlO_x$ stack for the long channel devices, ~60nm $HfO_2$/ $AlO_x$ for the short channel devices, limited by the bulging of gate dielectrics on the source/drain side wall), a clear upturn of DIBL is only observed at a channel length of 32nm. Extrapolating to a device with a 3nm $HfO_2$ gate insulator would predict a limiting channel length feature of ~7nm. Theoretically, for a MOSFET with monolayer $MoS_2$ (channel thickness: ~0.8nm, dielectric constant: 6.8~7.1 $\varepsilon_0$ , where $\varepsilon_0$ is the vacuum permittivity [61]) and 1nm equivalent oxide thickness (EOT), the electrostatic scaling length λ is only about 1.2nm.

These considerations suggest that MoS$_2$ could be a very promising material for scaled, high-density electronics.

**Breakdown electric fields**

The large band gap of MoS$_2$ implies the possibility of device operation at higher voltages or electric fields. Fig. 5b shows the critical breakdown fields of graphene and CVD MoS$_2$ MOSFETs devices. The measurement setup is shown on the upper-left inset of Fig. 5b. The channel-length dependence of breakdown voltage of MoS$_2$ MOSFETs is shown in the lower-right inset of Fig. 5b. The critical field can be extracted from the slope of the breakdown voltage vs channel length. Here we extracted the breakdown field from MoS$_2$ transistors with channel lengths of 80nm and 285nm. (More details about the breakdown test results of MoS$_2$ transistors and graphene transistors are shown in the Supplementary Figure S5 and S6, and Supplementary Note 2 and 3) The measured breakdown field of CVD MoS$_2$ is about 5 times larger than that of graphene, and significantly larger than that of SOI with 100nm silicon thickness [62]. In this regard, MoS$_2$ can also be a very promising platform for power devices.

**Discussion**

We have systematically studied the electronic transport properties of CVD MoS$_2$ devices. We report the observation of a significant amount of electronic trap states through capacitance and ac conductance measurements and their impact on the low-field electronic properties of MoS$_2$ devices. In particular, the measured effective mobility significantly underestimates the band mobility. An anomalous sub-threshold behavior, with distinctive temperature insensitivity, is also accounted for by the presence of these band tail states. We also studied the high-field electronic properties of $MoS_2$ devices and demonstrated the possibility to aggressively scale them down and their high breakdown electric fields. These attractive device attributes present a compelling case for wafer-scale monolayer $MoS_2$ as alternative to organic and other thin film materials for flexible electronics and photonics, including high resolution displays, photo-detection, logic electronics, power devices with solar energy harvesting etc. From the fundamental material standpoint, understanding of the microscopic origin of these band tail states is critical for further improvement of the material's electronic properties.

## Methods:

**Fabrication**

Large-scale monolayer $MoS_2$ was synthesized at 650 C by APCVD using perylene-3,4,9,10-tetracarboxylic acid tetra-potassium salt (PTAS) as the seed on $SiO_2$/Si substrate [39]. Sulfur powder and molybdenum oxide ($MoO_3$) were used as the precursors for the synthesis. The $SiO_2$ thickness was 300 nm. In the Hall-bar and transistor devices, the source/drain contact metal stack consisted of Ti/Au/Ti (5/15/5 nm). The $MoS_2$ channel was patterned using electron beam lithography and oxygen plasma etching. The top gate dielectric was comprised of an $AlO_x$/$HfO_2$ stack. The $AlO_x$ was formed by electron beam evaporation of 2 nm of aluminum metal followed by its natural oxidization in air for a few hours. The 30 nm thick $HfO_2$ layer was formed using atomic layer deposition (ALD) at 170 degrees. The top gate electrode was Ti/Au (5/40 nm).

**Characterization**

The capacitances were measured using Agilent B1500 Semiconductor Device Analyzer produced by Agilent technology. The temperature dependence of conductance was measured using cryogenic probestation produced in Lake Shore Cryotronics, Inc. The Raman spectrum was taken using Labram Aramis produced by Horiba Jobin Yvon. Scanning electron microscopy was measured using Leo 1560 produced by Carl Zeiss.

**Acknowledgement:**

We would like to thank Bruce Ek, Jim Bucchignano and Simon Dawes for their contributions to device fabrication. We would also like to thank Jin Cai and Vasili Perebeinos in IBM, Xiao Sun and Prof. Tso-Ping Ma at Yale University, and Prof. Mingfu Li at Fudan University for valuable discussions.

**Author contributions:**

W.Z. and F.X. initiated the project. Y-H.L. and J.K. carried out CVD $MoS_2$ growth. W.Z., F.X. and H.W. contributed to device design and fabrication. W.Z. performed the electrical characterization. W.Z., T.L, and F.X. analyzed the data. T.L. performed the modeling. D.B.F. grew the ALD gate dielectric. Y-H.L. carried out Atomic force spectroscopy (AFM). W.Z. performed Scanning electron spectroscopy (SEM) and Raman spectrum measurement. W.Z. designed, fabricated and measured the graphene devices. P.A. supervised this project. All authors participated in writing of the paper.

Competing financial interests: The authors declare no competing financial interests.

**References:**

1   Novoselov, K. S. *et al.* Two-dimensional atomic crystals. *Proceedings of the National Academy of Sciences of the United States of America* **102**, 10451-10453 (2005).
2   Chhowalla, M. *et al.* The chemistry of two-dimensional layered transition metal dichalcogenide nanosheets. *Nature Chemistry* **5**, 263-275 (2013).
3   Hsu, A. *et al.* Large-area 2-D electronics: materials, technology, and devices. *Proceedings of the IEEE* **101**, 1638-1652 (2013).
4   Mak, K. F., Lee, C., Hone, J., Shan, J. & Heinz, T. F. Atomically Thin $MoS_{2}$: A New Direct-Gap Semiconductor. *Physical Review Letters* **105**, 136805 (2010).
5   Sundaram, R. S. *et al.* Electroluminescence in Single Layer MoS2. *Nano Letters* **13**, 1416-1421 (2013).
6   Lee, H. S. *et al.* MoS2 Nanosheet Phototransistors with Thickness-Modulated Optical Energy Gap. *Nano Letters* **12**, 3695-3700 (2012).
7   Yin, Z. *et al.* Single-Layer MoS2 Phototransistors. *Acs Nano* **6**, 74-80 (2011).
8   Dashora, A., Ahuja, U. & Venugopalan, K. Electronic and optical properties of MoS2 thin films: Feasibility for solar cells. *Computational Materials Science* **69**, 216-221 (2013).
9   Pu, J. *et al.* Highly Flexible MoS2 Thin-Film Transistors with Ion Gel Dielectrics. *Nano Letters* **12**, 4013-4017 (2012).
10  Chang, H.-Y. *et al.* High-Performance, Highly Bendable MoS2 Transistors with High-K Dielectrics for Flexible Low-Power Systems. *Acs Nano* **7**, 5446-5452 (2013).
11  Ayari, A., Cobas, E., Ogundadegbe, O. & Fuhrer, M. S. Realization and electrical characterization of ultrathin crystals of layered transition-metal dichalcogenides. *Journal of Applied Physics* **101**, 014507-014505 (2007).
12  Benameur, M. M. *et al.* Visibility of dichalcogenide nanolayers. *Nanotechnology* **22** (2011).
13  Buscema, M. *et al.* Large and Tunable Photothermoelectric Effect in Single-Layer MoS2. *Nano Letters* **13**, 358-363 (2013).
14  Ghatak, S., Pal, A. N. & Ghosh, A. Nature of Electronic States in Atomically Thin MoS2 Field-Effect Transistors. *Acs Nano* **5**, 7707-7712 (2011).
15  Kim, J.-Y., Choi, S. M., Seo, W.-S. & Cho, W.-S. Thermal and Electronic Properties of Exfoliated Metal Chalcogenides. *Bulletin of the Korean Chemical Society* **31**, 3225-3227 (2010).
16  Lembke, D. & Kis, A. Breakdown of High-Performance Monolayer MoS2 Transistors. *Acs Nano* **6**, 10070-10075 (2012).
17  Radisavljevic, B. & Kis, A. Mobility engineering and a metal–insulator transition in monolayer MoS2. *Nat Mater* advance online publication (2013).
18  RadisavljevicB, RadenovicA, BrivioJ, GiacomettiV & KisA. Single-layer MoS2 transistors. *Nat Nano* **6**, 147-150 (2011).
19  Liu, H., Neal, A. T. & Ye, P. D. Channel Length Scaling of MoS2 MOSFETs. *Acs Nano* **6**, 8563-8569 (2012).

20	Kim, S. *et al.* High-mobility and low-power thin-film transistors based on multilayer MoS2 crystals. *Nature Communications* 3 (2012).
21	Bao, W., Cai, X., Kim, D., Sridhara, K. & Fuhrer, M. S. High mobility ambipolar MoS2 field-effect transistors: Substrate and dielectric effects. *Applied Physics Letters* 102 (2013).
22	Jariwala, D. *et al.* Band-like transport in high mobility unencapsulated single-layer MoS[sub 2] transistors. *Applied Physics Letters* 102, 173107 (2013).
23	Li, S.-L. *et al.* Thickness-Dependent Interfacial Coulomb Scattering in Atomically Thin Field-Effect Transistors. *Nano Letters* (2013).
24	Perera, M. M. *et al.* Improved Carrier Mobility in Few-Layer MoS2 Field-Effect Transistors with Ionic-Liquid Gating. *Acs Nano* 7, 4449-4458 (2013).
25	Pradhan, N. R. *et al.* Intrinsic carrier mobility of multi-layered MoS[sub 2] field-effect transistors on SiO[sub 2]. *Applied Physics Letters* 102, 123105 (2013).
26	Zhang, Y., Ye, J., Matsuhashi, Y. & Iwasa, Y. Ambipolar MoS2 Thin Flake Transistors. *Nano Letters* 12, 1136-1140 (2012).
27	Qiu, H. *et al.* Electrical characterization of back-gated bi-layer MoS[sub 2] field-effect transistors and the effect of ambient on their performances. *Applied Physics Letters* 100, 123104 (2012).
28	Das, S., Chen, H.-Y., Penumatcha, A. V. & Appenzeller, J. High Performance Multilayer MoS2 Transistors with Scandium Contacts. *Nano Letters* 13, 100-105 (2013).
29	Liu, K.-K. *et al.* Growth of Large-Area and Highly Crystalline MoS2 Thin Layers on Insulating Substrates. *Nano Letters* 12, 1538-1544 (2012).
30	Najmaei, S. *et al.* Vapour phase growth and grain boundary structure of molybdenum disulphide atomic layers. *Nat Mater* 12, 754-759 (2013).
31	Lee, Y.-H. *et al.* Synthesis of Large-Area MoS2 Atomic Layers with Chemical Vapor Deposition. *Advanced Materials* 24, 2320-2325 (2012).
32	Zhan, Y., Liu, Z., Najmaei, S., Ajayan, P. M. & Lou, J. Large-Area Vapor-Phase Growth and Characterization of MoS2 Atomic Layers on a SiO2 Substrate. *Small* 8, 966-971 (2012).
33	Wang, H. *et al.* in *Electron Devices Meeting (IEDM), 2012 IEEE International.*  4.6.1-4.6.4.
34	Amani, M. *et al.* Electrical performance of monolayer MoS[sub 2] field-effect transistors prepared by chemical vapor deposition. *Applied Physics Letters* 102, 193107 (2013).
35	Liu, H. *et al.* Statistical Study of Deep Submicron Dual-Gated Field-Effect Transistors on Monolayer Chemical Vapor Deposition Molybdenum Disulfide Films. *Nano Letters* 13, 2640-2646 (2013).
36	Wu, W. *et al.* High mobility and high on/off ratio field-effect transistors based on chemical vapor deposited single-crystal MoS[sub 2] grains. *Applied Physics Letters* 102, 142106 (2013).
37	Zhou, W. *et al.* Intrinsic Structural Defects in Monolayer Molybdenum Disulfide. *Nano Letters* 13, 2615-2622 (2013).

| 38 | Fuhrer, M. S. & Hone, J. Measurement of mobility in dual-gated MoS2 transistors. *Nat Nano* **8**, 146-147 (2013). |
|---|---|
| 39 | Lee, Y.-H. *et al.* Synthesis and Transfer of Single-Layer Transition Metal Disulfides on Diverse Surfaces. *Nano Letters* **13**, 1852-1857 (2013). |
| 40 | Li, S.-L. *et al.* Quantitative Raman Spectrum and Reliable Thickness Identification for Atomic Layers on Insulating Substrates. *Acs Nano* **6**, 7381-7388 (2012). |
| 41 | Kaasbjerg, K., Thygesen, K. S. & Jacobsen, K. W. Phonon-limited mobility in n-type single-layer MoS2 from first principles. *Physical Review B* **85** (2012). |
| 42 | Pollak, M. & Shklovskii, B. *Hopping Transport in Solids*. Vol. 28 (Elsevier Science Publishers B.V., 1991). |
| 43 | Qiu, H. *et al.* Hopping transport through defect-induced localized states in molybdenum disulphide. *Nat Commun* **4** (2013). |
| 44 | Nicollian, E. H. & Brews, J. R. *MOS (metal oxide semiconductor) physics and technology*. (Wiley-Interscience Publication, 1982). |
| 45 | Shi, H., Pan, H., Zhang, Y.-W. & Yakobson, B. I. Quasiparticle band structures and optical properties of strained monolayer MoS_{2} and WS_{2}. *Physical Review B* **87**, 155304 (2013). |
| 46 | Lebègue, S. & Eriksson, O. Electronic structure of two-dimensional crystals from ab initio theory. *Physical Review B* **79**, 115409 (2009). |
| 47 | Mott, N. F. & Davis, E. A. *Electronic processes in non-crystalline materials*. (Oxford Press, 1979). |
| 48 | Mott, N. F. Conduction in non-crystalline materials. *Philosophical Magazine* **21**, 863-867 (1969). |
| 49 | Monroe, D. Hopping in Exponential Band Tails. *Physical Review Letters* **54**, 146-149 (1985). |
| 50 | Bässler, M. S. H. Calculation of energy relaxation and transit time due to hopping in an exponential band tail. *Phil. Mag. Lett.* **56** (1987). |
| 51 | Tiedje, T. & Rose, A. A physical interpretation of dispersive transport in disordered semiconductors. *Solid State Communications* **37**, 49-52 (1981). |
| 52 | Hori, T. *Gate dielectrics and MOS ULSIs principles, technologies, and applications*. (Springer, 1997). |
| 53 | Salleo, A. *et al.* Intrinsic hole mobility and trapping in a regioregular poly(thiophene). *Physical Review B* **70**, 115311 (2004). |
| 54 | Völkel, A. R., Street, R. A. & Knipp, D. Carrier transport and density of state distributions in pentacene transistors. *Physical Review B* **66**, 195336 (2002). |
| 55 | Ferry, D. *Transport in Nanostructure*. Vol. Ch.2 (Cambridge University Press, 2009). |
| 56 | Yuan Taur, T. H. N. *Fundamentals of Modern VLSI Devices*. (Cambridge University Press, 1998). |
| 57 | Ran-Hong, Y., Ourmazd, A. & Lee, K. F. Scaling the Si MOSFET: from bulk to SOI to bulk. *Electron Devices, IEEE Transactions on* **39**, 1704-1710 (1992). |

58	Majumdar, A., Zhibin, R., Koester, S. J. & Haensch, W. Undoped-Body Extremely Thin SOI MOSFETs With Back Gates. *Electron Devices, IEEE Transactions on* **56**, 2270-2276 (2009).
59	Low, T. *et al.* Modeling study of the impact of surface roughness on silicon and Germanium UTB MOSFETs. *Electron Devices, IEEE Transactions on* **52**, 2430-2439 (2005).
60	Ohashi, T., Takahashi, T., Beppu, N., Oda, S. & Uchida, K. in *Electron Devices Meeting (IEDM), 2011 IEEE International.* 16.14.11-16.14.14.
61	Salmani-Jelodar, M., Yaohua, T. & Klimeck, G. in *Semiconductor Device Research Symposium (ISDRS), 2011 International.* 1-2.
62	Merchant, S. *et al.* in *Power Semiconductor Devices and ICs, 1991. ISPSD '91., Proceedings of the 3rd International Symposium on.* 31-35.

**Figure legends:**

Figure 1. **Chemical vapor deposition (CVD) molybdenum disulphide (MoS$_2$) physical properties and device structure.** (a) Schematic of MOSFET with monolayer CVD grown MoS$_2$. (b) AFM image of CVD grown MoS$_2$ on SiO$_2$/Si substrate. The scale bar in the AFM image is 10μm. The inset shows the step height profile of MoS$_2$ in the AFM image.

Figure 2. **Characterization and modeling of the density and dynamics of band tail states.** (a) Capacitance as a function of frequency measured at various gate voltages. The device width is 4μm and length is 44μm. The symbols are the experimental results and the lines are fittings using the device model shown in (b). (b) The equivalent circuit model of the device, simplified parallel model and the measurement model in series mode. (c) Density and time constant of trap states as a function of gate voltages. The symbols are experimental results extracted from the capacitance $C_{ms}$ and ac conductance $G_p$. The lines are models, see text. (d) Extracted ac conductance over angular frequency $G_p/\omega$ as a function of frequency $f$ at various gate voltages. (e). Parameterized model, $D_n(E)$, describing the electronic density-of-states of both the extended and localized states. The valence density-of-state is also included in the illustration using the mirror image of conduction density-of-state. The inset illustrates the band diagram of MoS$_2$ MOSFET. (f) Multi-frequency capacitance of MoS$_2$ Hall-bar as a function of gate voltage. The symbols are experimental results and the lines are the modeling results.

Figure 3. **Temperature dependence of conductance and activation energy.** (a) Four point conductance as a function of gate voltage measured at various temperatures from 4.4K to 400K on a MoS$_2$ Hall-bar. (b) Conductance as a function of reverse of temperature 1/T at various gate voltages. The inset shows the activation energy at various gate voltages fitted to the Arrhenius activated energy model. The symbols are experimental results extracted from the conductance, the solid line gives the modeling results. (c) The calculated band carriers as a function of gate voltage at various temperatures.

Figure 4. **Effective mobility and band mobility.** (a) Effective mobility as a function of gate voltage measured at various temperatures in MoS$_2$ Hall-bar. (b) Effective mobility and the corresponding band mobility as a function of temperature at $V_{TG} = 4V$. The mobility limited by MoS$_2$ phonons scattering based on theoretical calculations[41] are also plotted for comparison.

Figure 5. **Device properties at high electrical fields.** (a) Drain-induced-barrier-lowering (DIBL) of MOSFET with CVD MoS$_2$ at various channel lengths. The inset shows the SEM image of the 32nm and 80nm channel length devices. (b) Critical electric field of CVD MoS$_2$ and graphene. The upper inset shows the measurement configuration. The lower inset shows the lateral breakdown voltage as a function of channel length for MOSFET with CVD MoS$_2$.

a

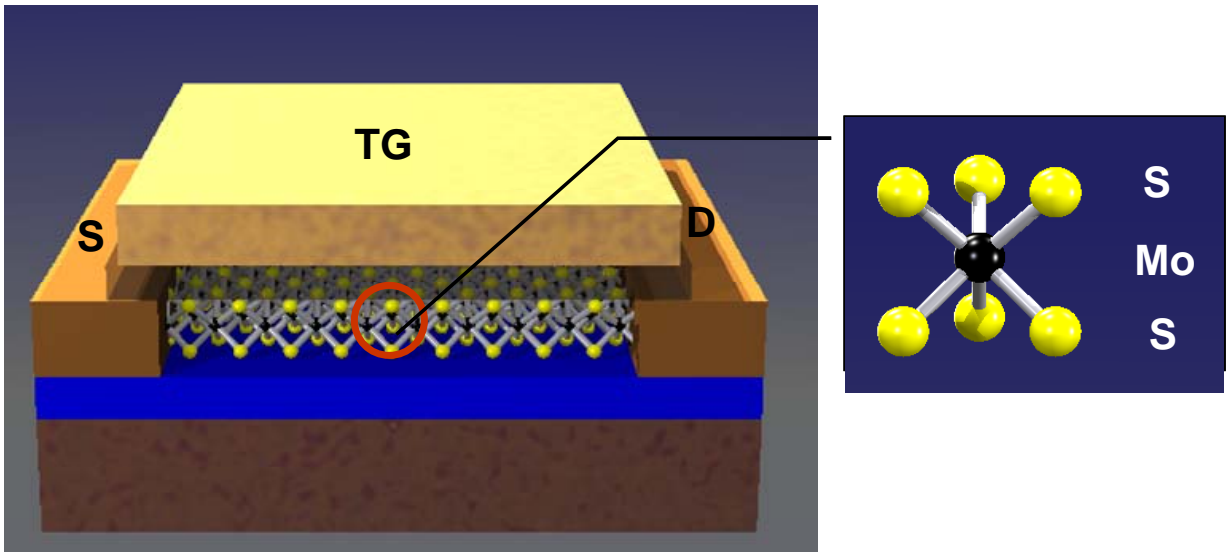

b

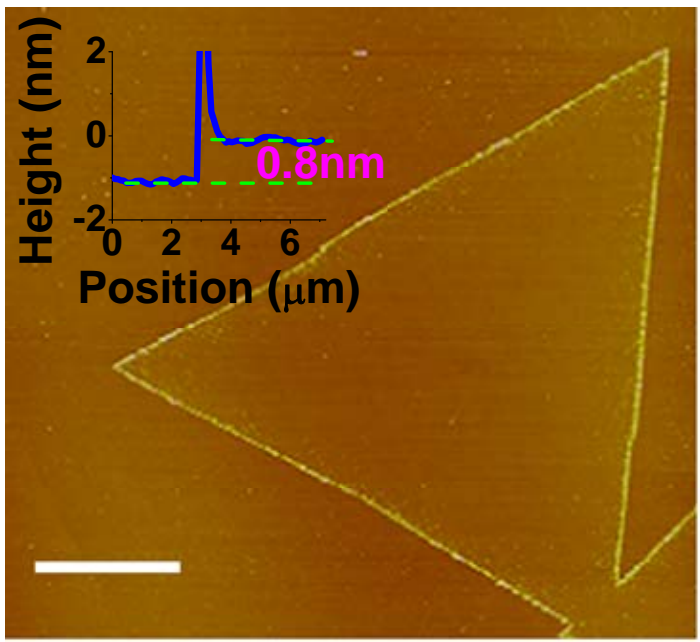

**Figure-1 (Zhu)**

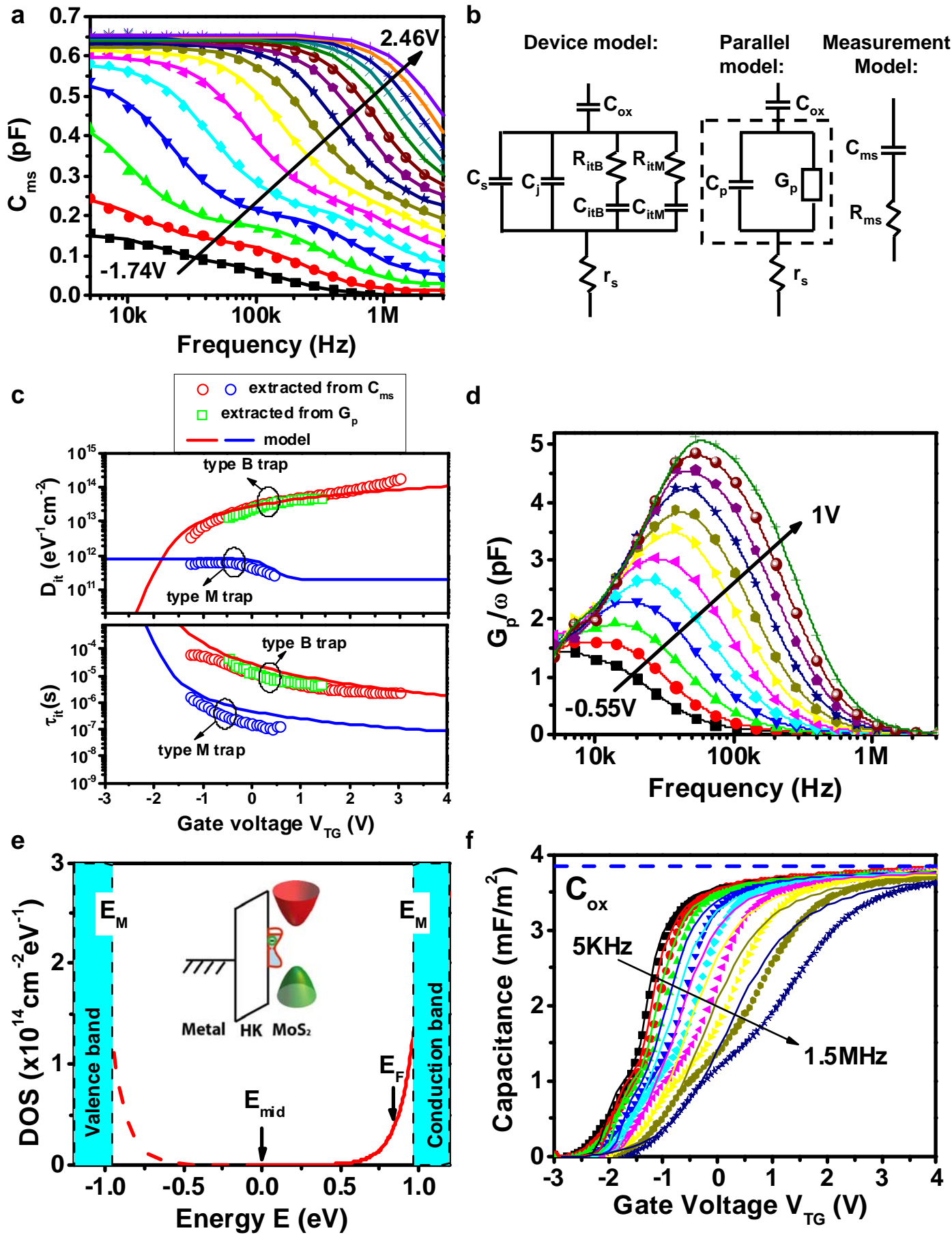

Figure-2 (Zhu)

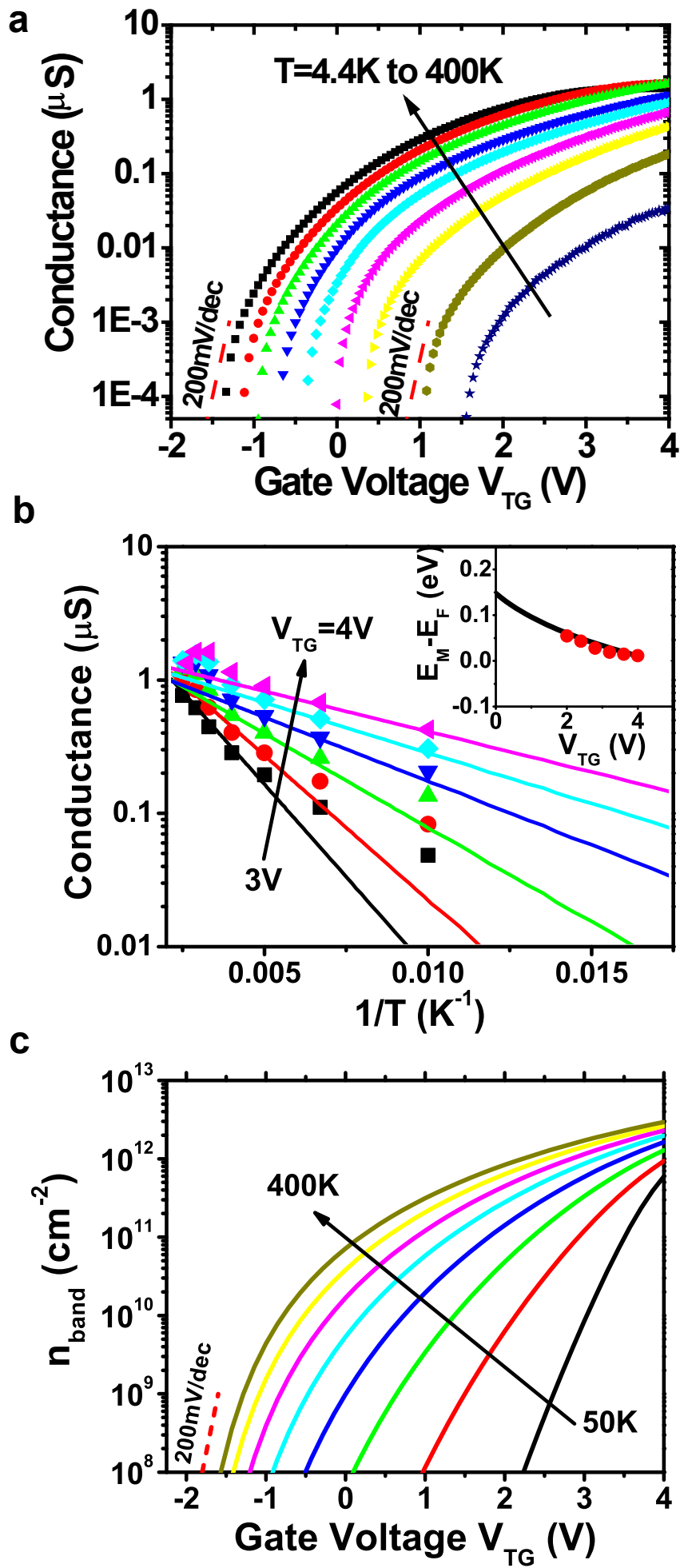

Figure-3 (Zhu)

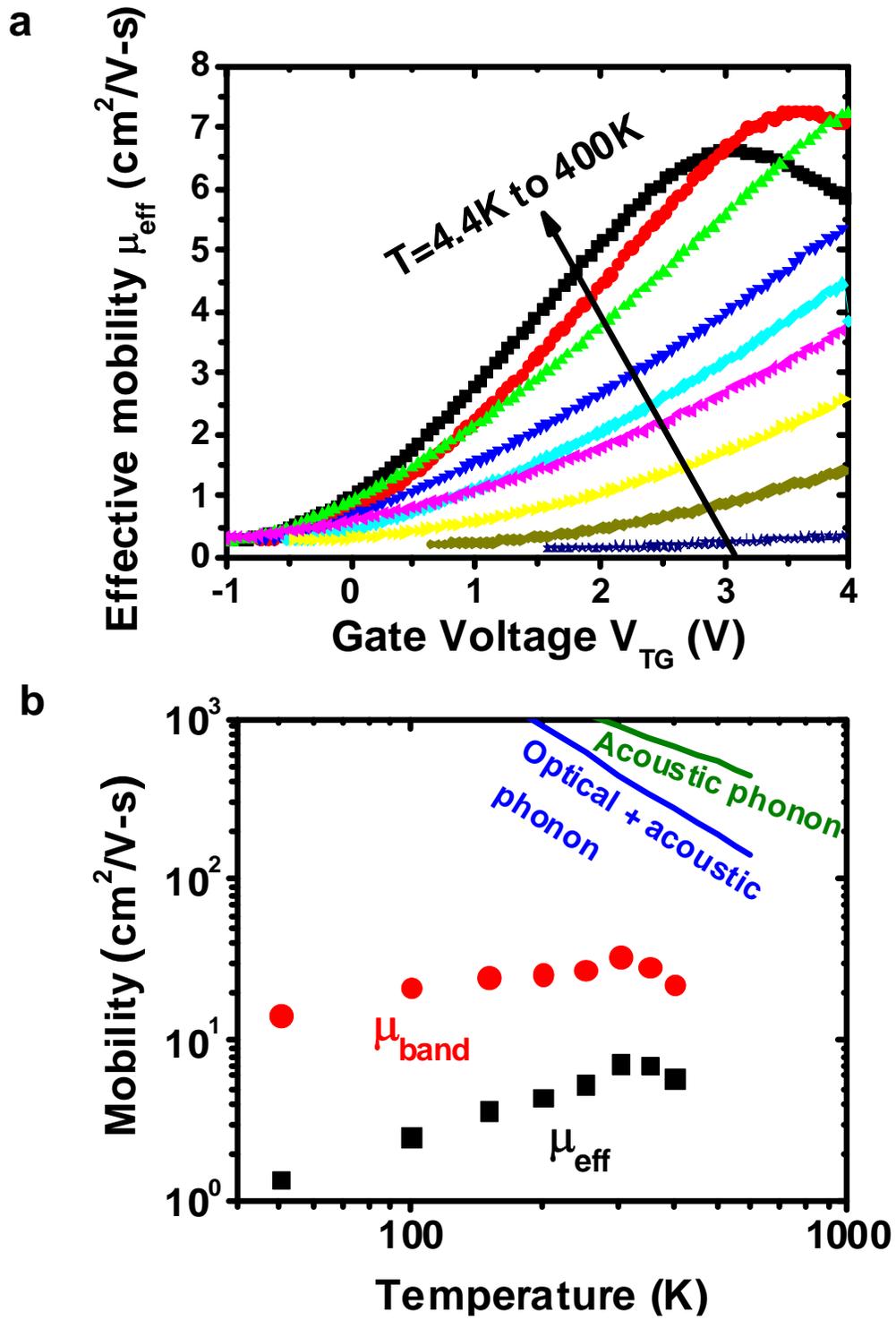

**Figure-4 (Zhu)**

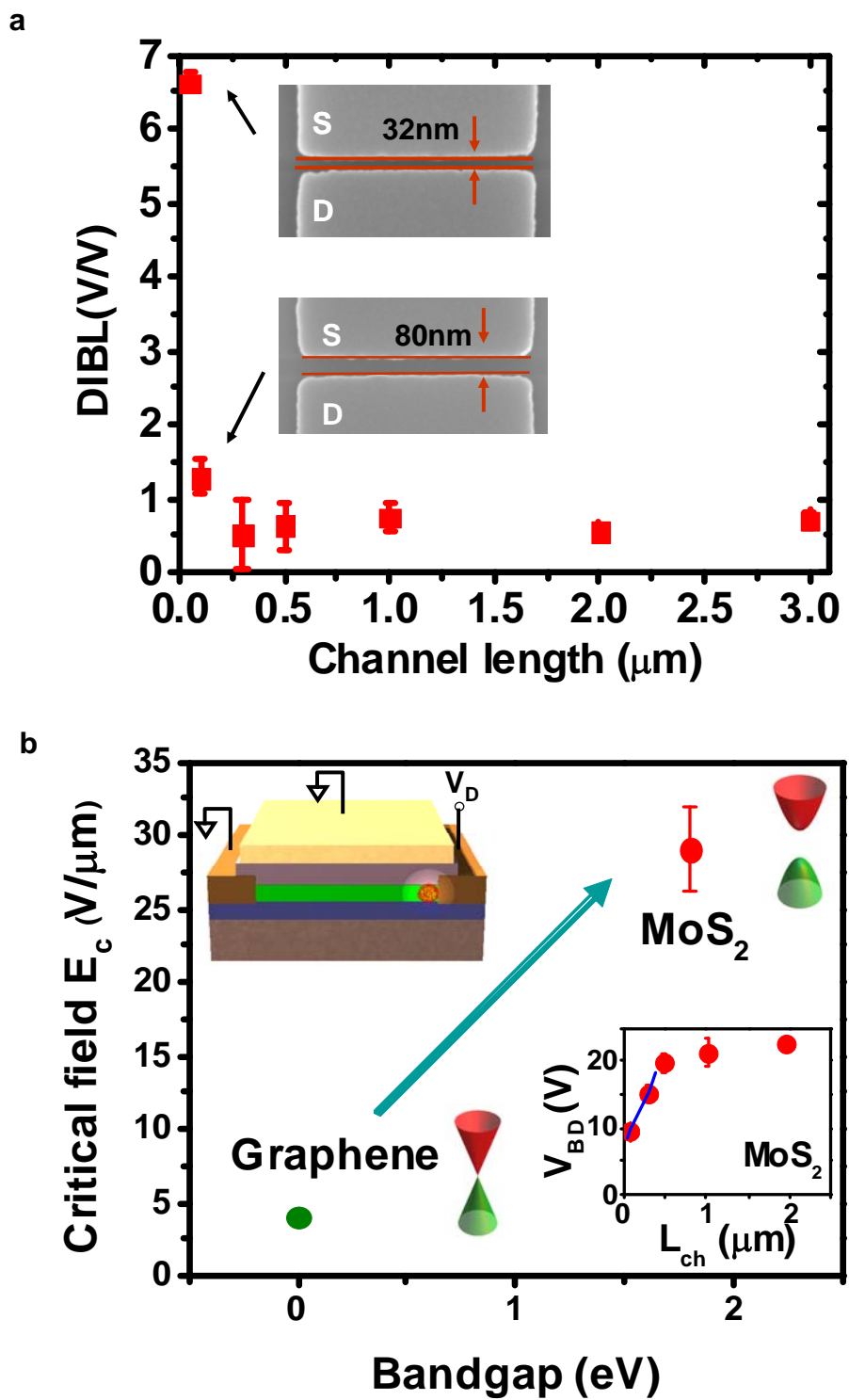

**Figure-5 (Zhu)**